# Properties of Phase Space Wavefunctions and Eigenvalue Equation of Momentum Dispersion Operator

Ravo Tokiniaina Ranaivoson[1], Raoelina Andriambololona[2], Hanitriarivo Rakotoson[3]

*raoelinasp@yahoo.fr[1];jacquelineraoelina@hotmail.com[1];raoelina.andriambololona@gmail.com[1]; tokhiniaina@gmail.com[2],infotsara@gmail.com[3]*

*Theoretical Physics Department*

*Institut National des Sciences et Techniques Nucléaires (INSTN- Madagascar)*

BP 4279 Antananarivo 101, Madagascar, *instn@moov.mg*

***Abstract***: This paper is a continuation of our previous works about coordinate, momentum, dispersion operators and phase space representation of quantum mechanics. It concerns a study on the properties of wavefunctions in the phase space representation and the momentum dispersion operator, its representations and eigenvalue equation. After the recall of some results from our previous papers, we give most of the main properties of the phase space wavefunctions and consider some examples of them. Then we establish the eigenvalue equation for the differential operator corresponding to the momentum dispersion operator in the phase space representation. It is shown in particular that any phase space wavefunction is solution of this equation.

## 1-Introduction

The present work can be considered as part of a series of studies related to a phase space representation of quantum theory introduced and developed in [1], [2] and [3]. Because of the uncertainty relation [4], the problem of considering phase space, which mix momentum with coordinate, in quantum theory is an interesting challenge. Many works related to this subject have been already performed. We may quote for instance [5-12]. Many of these works are based on the approach introduced by Wigner in [5].

In the reference [1], we have considered an approach using the current formulation of quantum mechanics based on linear operators theory and Hilbert space (as in [4] for instance). Our previous work [1] may be considered as extension in the framework of quantum theory of the results obtained in [14]. The phase space representation that we have defined is based on the introduction of quantum states, denoted $|n, X, P, \mathcal{B}\rangle$. These states are defined by the means values $X, P$ and statistical dispersions (variance) $(\Delta x_n)^2$ and $(\Delta p_n)^2$ of coordinate and momentum ($n$ being a positive integer number). We have the relations

$$\begin{cases}(\Delta x_n)^2 = (2n+1)(\mathcal{a})^2 = (2n+1)\mathcal{A} \\ (\Delta p_n)^2 = (2n+1)(\mathcal{b})^2 = (2n+1)\mathcal{B} \\ \mathcal{A} = (\mathcal{a})^2 \quad \mathcal{B} = (\mathcal{b})^2 \\ \mathcal{a}\mathcal{b} = \dfrac{\hbar}{2}\end{cases} \tag{1}$$

$$\begin{cases}\langle n, X, P, \mathcal{b}|\boldsymbol{x}|n, X, P, \mathcal{b}\rangle = X \\ \langle n, X, P, \mathcal{b}|\boldsymbol{p}|n, X, P, \mathcal{b}\rangle = P \\ \left\langle n, X, P, \mathcal{b}\middle|(\boldsymbol{x}-X)^2\middle|n, X, P, \mathcal{b}\right\rangle = (2n+1)(\mathcal{a})^2 \\ \left\langle n, X, P, \mathcal{b}\middle|(\boldsymbol{p}-P)^2\middle|n, X, P, \mathcal{b}\right\rangle = (2n+1)(\mathcal{b})^2\end{cases} \tag{2}$$

$\boldsymbol{p}$ and $\boldsymbol{x}$ being respectively the momentum and coordinate operators. The wavefunctions corresponding to a state $|n, X, P, \mathcal{b}\rangle$ respectively in coordinate and momentum representations are the Harmonic Hermite-Gaussian functions $\varphi_n$ and their Fourier transforms $\tilde{\varphi}_n$ defined in [1] and [14].

$$\varphi_n(x, X, P, \mathcal{b}) = \langle x|n, X, P, \mathcal{b}\rangle = \frac{H_n\left(\frac{x-X}{\sqrt{2}\mathcal{a}}\right)}{\sqrt{2^n n!\sqrt{2\pi}\mathcal{a}}} e^{-\left(\frac{x-X}{2\mathcal{a}}\right)^2 + i\frac{Px}{\hbar}} \tag{3}$$

$$\tilde{\varphi}_n(p, X, P, \mathcal{b}) = \langle p|n, X, P, \mathcal{b}\rangle = \frac{(-i)^n H_n\left(\frac{p-P}{\sqrt{2}\mathcal{b}}\right)}{\sqrt{2^n n!\sqrt{2\pi}\mathcal{b}}} e^{-\left(\frac{p-P}{2\mathcal{b}}\right)^2 - iX\frac{(p-P)}{\hbar}} \tag{4}$$

$$\tilde{\varphi}_n(p, X, P, \mathcal{b}) = \frac{1}{\sqrt{2\pi\hbar}} \int \varphi_n(x, X, P, \mathcal{b})\, e^{-i\frac{px}{\hbar}} dx$$

$H_n$ is a Hermite polynomial of order $n$. It has been established that a state $|n, X, P, \mathcal{b}\rangle$ is an eigenstate of the momentum dispersion operator $\boldsymbol{\Sigma}_{\boldsymbol{p}} = \beth^+$ and the coordinate dispersion operator $\boldsymbol{\Sigma}_{\boldsymbol{x}} = \frac{\mathcal{A}}{\mathcal{B}}\beth^+$. The explicit expressions of these operators are

$$\begin{cases}\boldsymbol{\Sigma}_{\boldsymbol{p}} = \beth^+ = \dfrac{1}{2}[(\boldsymbol{p}-P)^2 + 4\mathcal{B}(\boldsymbol{x}-X)^2] \\ \boldsymbol{\Sigma}_{\boldsymbol{x}} = \dfrac{\mathcal{A}}{\mathcal{B}}\beth^+ = \dfrac{1}{2}[4\mathcal{A}(\boldsymbol{p}-P)^2 + (\boldsymbol{x}-X)^2]\end{cases} \tag{5}$$

and the corresponding eigenvalues equations are

$$\begin{cases}\boldsymbol{\Sigma}_{\boldsymbol{p}}|n, X, P, \mathcal{b}\rangle = (2n+1)\mathcal{B}|n, X, P, \mathcal{b}\rangle \\ \boldsymbol{\Sigma}_{\boldsymbol{x}}|n, X, P, \mathcal{b}\rangle = (2n+1)\mathcal{A}|n, X, P, \mathcal{b}\rangle\end{cases} \tag{6}$$

As the eigenvalues of the momentum and coordinate dispersion operators are proportional and as they have the same eigenstates, it is sufficient to consider only the momentum dispersion operator $\beth^+$.

In the reference [2], we have tackled the problem of finding the representations of coordinate, momentum and dispersion operators in the framework of the phase space representation. We have established that they can be at the same time represented both with matrix and differential operators in the basis $\{|n, X, P, \mathcal{b}\rangle\}$ defining this phase space representation.

In this paper, our goal is to perform a study on the properties of wavefunctions in the phase space representation and their relation with the eigenvalue equation of the momentum dispersion operator $\beth^+$. We show and verify explicitly, in particular, that these wavefunctions are as expected the eigenfuctions of the differential operator which represents $\beth^+$.

## 2-Phase space representations of momentum dispersion operator

Let us consider the momentum dispersion operator $\beth^+$. It can be put in the form

$$\beth^+ = \frac{1}{2}[\,(\boldsymbol{p}-P)^2 + 4\frac{\mathcal{B}^2}{\hbar^2}(\boldsymbol{x}-X)^2] = 4\mathcal{B}\underline{\beth}^+ = \mathcal{B}(\mathfrak{p}^2 + \mathfrak{x}^2) \tag{7}$$

in which

$$\begin{cases} \underline{\beth}^+ = \dfrac{1}{4}(\mathfrak{p}^2 + \mathfrak{x}^2) \\ \mathfrak{p} = \dfrac{(\boldsymbol{p}-P)}{\sqrt{2}\mathscr{b}} = \dfrac{(\boldsymbol{p}-P)}{\sqrt{2\mathcal{B}}} = \dfrac{\sqrt{2}\mathscr{a}}{\hbar}(\boldsymbol{p}-P) = \dfrac{\sqrt{2\mathcal{A}}}{\hbar}(\boldsymbol{p}-P) \\ \mathfrak{x} = \dfrac{(\boldsymbol{x}-X)}{\sqrt{2}\mathscr{a}} = \dfrac{(\boldsymbol{x}-X)}{\sqrt{2\mathcal{A}}} = \dfrac{\sqrt{2}}{\hbar}\mathscr{b}(\boldsymbol{x}-X) = \dfrac{\sqrt{2\mathcal{B}}}{\hbar}(\boldsymbol{x}-X) \end{cases} \tag{8}$$

In the paper [2], we have established that in the phase space representation, we have for the operators $\mathfrak{p}$ and $\mathfrak{x}$ on one hand the matrix representations

$$\begin{cases} \mathfrak{p}_m^n = \langle n, X, P, \mathscr{b}|\mathfrak{p}|m, X, P, \mathscr{b}\rangle = \dfrac{1}{\sqrt{2}}(\sqrt{m}\delta_{m-1}^n + \sqrt{m+1}\delta_{m+1}^n) \\ \mathfrak{x}_m^n = \langle n, X, P, \mathscr{b}|\mathfrak{x}|m, X, P, \mathscr{b}\rangle = \dfrac{i}{\sqrt{2}}(\sqrt{m}\delta_{m-1}^n - \sqrt{m+1}\delta_{m+1}^n) \end{cases} \tag{9}$$

which satisfy the commutation relation

$$\mathfrak{x}_l^n\mathfrak{p}_m^l - \mathfrak{p}_l^n\mathfrak{x}_m^l = i\delta_m^n \tag{10}$$

and on the other hand the differential operators representations:

$$\begin{cases} \hat{\mathfrak{p}} = \dfrac{1}{\sqrt{2}\mathscr{a}}(i\hbar\dfrac{\partial}{\partial P} - X) = \sqrt{2}\mathscr{b}(i\dfrac{\partial}{\partial P} - \dfrac{X}{\hbar}) \\ \hat{\mathfrak{x}} = -i\sqrt{2}\mathscr{a}\dfrac{\partial}{\partial X} = -i\dfrac{\hbar}{\sqrt{2}\mathscr{b}}\dfrac{\partial}{\partial X} \end{cases} \tag{11}$$

From the relation (8) and (9),we obtain for the matrix representation of the operator $\underline{\beth}^+$

$$\underline{\beth}^{+}{}_m^n = \langle n, X, P, \mathscr{b}|\underline{\beth}^+|m, X, P, \mathscr{b}\rangle = \frac{1}{4}(\mathfrak{p}_l^n\mathfrak{p}_m^l + \mathfrak{x}_l^n\mathfrak{x}_m^l) = \frac{1}{4}(2n+1\,)\delta_m^n \tag{12}$$

and from the relation (8) and (11), we obtain for its differential operator representation

$$\hat{\underline{\beth}}^+ = \frac{1}{4}(\hat{\mathfrak{p}}^2 + \hat{\mathfrak{x}}^2) = \frac{1}{2}[-\mathcal{A}\frac{\partial^2}{\partial X^2} - \mathcal{B}\frac{\partial^2}{\partial P^2} - 2i\frac{\mathcal{B}}{\hbar}X\frac{\partial}{\partial P} + \frac{\mathcal{B}}{\hbar^2}X^2] \tag{13}$$

Using the relation (7), (12), (13) and (1), we obtain respectively for the matrix and differential operator representations of the momentum dispersion operator $\beth^{+}$

$$\beth^{+n}_{\ m} = \langle n,X,P,\mathscr{b}|\beth^{+}|m,X,P,\mathscr{b}\rangle = 4\mathcal{B}\underline{\beth}^{+n}_{\ m} = (2n+1\,)\mathcal{B}\delta^{n}_{m} \tag{14}$$

$$\tilde{\beth}^{+} = 4\mathcal{B}\,\tilde{\underline{\beth}}^{+} = [-\frac{\hbar^{2}}{2\mathcal{B}}\frac{\partial^{2}}{\partial X^{2}} - 2\mathcal{B}\frac{\partial^{2}}{\partial P^{2}} - \frac{4i\mathcal{B}}{\hbar}X\frac{\partial}{\partial P} + \frac{2\mathcal{B}}{\hbar^{2}}X^{2}] \tag{15}$$

We may remark that the expression of $\beth^{+n}_{\ m}$ corresponds to the fact that the elements of the basis $\{|n,X,P,\mathscr{b}\rangle\}$ defining the phase space representation are the eigenvectors of $\beth^{+}$.

## 3-Phase space wavefunctions

A phase space wavefunction $\Psi^{n}(X,P,\mathscr{b})$ of a particle is a phase space representation of its quantum vector state $|\psi\rangle$. It is equal to the inner product of the vector $|\psi\rangle$ with an element of the basis $\{|n,X,P,\mathscr{b}\rangle\}$ defining the phase space representation. As established in our papers [1], we have explicitly the relations

$$\begin{aligned}\Psi^{n}(X,P,\mathscr{b}) = \langle n,X,P,\mathscr{b}|\psi\rangle &= \int_{-\infty}^{+\infty}\varphi_{n}^{*}(x,X,P,\mathscr{b})\psi(x)dx \\ &= \int_{-\infty}^{+\infty}\tilde{\varphi}_{n}^{*}(p,X,P,\mathscr{b})\tilde{\psi}(p)dp\end{aligned} \tag{16}$$

in which, the functions $\psi(x) = \langle x|\psi\rangle$ and $\tilde{\psi}(p) = \langle p|\psi\rangle$ are the wavefunctions corresponding to the state $|\psi\rangle$ respectively in coordinate and momentum representations. The functions $\varphi_{n}^{*}(x,X,P,\mathscr{b})$ and $\tilde{\varphi}_{n}^{*}(p,X,P,\mathscr{b})$ are the complex conjugates of the wavefunctions $\varphi_{n}(x,X,P,\mathscr{b})$ and $\tilde{\varphi}_{n}(p,X,P,\mathscr{b})$ corresponding to the state $|n,X,P,\mathscr{b}\rangle$ respectively in coordinate and momentum representations. The expressions of $\varphi_{n}$ and $\tilde{\varphi}_{n}$ are given in (3) and (4). There are two possibilities to expand a state $|\psi\rangle$ in the basis $\{|n,X,P,\mathscr{b}\rangle\}$ of the phase space representation. The first one corresponds to any fixed value of $X,P$ and $\mathscr{b}$ and the expansion is obtained by varying the positive integer $n$ :

$$|\psi\rangle = \sum_{n=0}^{+\infty}\Psi^{n}(X,P,\mathscr{b})\,|n,X,P,\mathscr{b}\rangle \tag{17}$$

The second one corresponds to any fixed value of $n$ and $\mathscr{b}$ and the expansion is obtained by varying the real number $X$ and $P$

$$|\psi\rangle = \int_{-\infty}^{+\infty}\int_{-\infty}^{+\infty}\Psi^{n}(X,P,\mathscr{b})|n,X,P,\mathscr{b}\rangle\frac{dXdP}{2\pi\hbar} \tag{18}$$

We can also have two anothers kind of expansion of a the state $|\psi\rangle$ : the first one is obtained by fixing the values of $n$ and $X$ and varying the values of the real number $P$ and the positive real number $\mathscr{b}$

$$|\psi\rangle = \int_{-\infty}^{+\infty}\int_{0}^{+\infty}\Psi^{n}(X,P,\mathscr{b})\,[(\boldsymbol{x}-X)|n,X,P,\mathscr{b}\rangle]\frac{dPd\mathscr{b}}{\pi\hbar\mathscr{b}} \tag{19}$$

the second one is obtained by fixing the values of $n$ and $P$ and varying the values of the real number $X$ and the positive real number $a = \frac{\hbar}{2\mathscr{b}}$

$$|\psi\rangle = (-1)^n \int_{-\infty}^{+\infty} \int_{0}^{+\infty} \Psi^n(X, P, \mathscr{b}) \, [(\boldsymbol{p} - P)|n, X, P, \mathscr{b}\rangle] \frac{da dX}{\pi \hbar a} \tag{20}$$

To the four above relations (17), (18), (19) and (20) correspond the following relations between $\psi(x), \tilde{\psi}(p), \varphi_n(x, X, P, \mathscr{b})$ and $\tilde{\varphi}_n(p, X, P, \mathscr{b})$ :

$$\psi(x) = \sum_{n=0}^{+\infty} \Psi^n(X, P, \mathscr{b}) \, \varphi_n(x, X, P, \mathscr{b})$$

$$\tilde{\psi}(p) = \sum_{n=0}^{+\infty} \Psi^n(X, P, \mathscr{b}) \, \tilde{\varphi}_n(p, X, P, \mathscr{b})$$

$$\psi(x) = \int_{-\infty}^{+\infty} \int_{-\infty}^{+\infty} \Psi^n(X, P, \mathscr{b}) \varphi_n(x, X, P, \mathscr{b}) \frac{dX dP}{2\pi\hbar}$$

$$\tilde{\psi}(p) = \int_{-\infty}^{+\infty} \int_{-\infty}^{+\infty} \Psi^n(X, P, \mathscr{b}) \tilde{\varphi}_n(p, X, P, \mathscr{b}) \frac{dX dP}{2\pi\hbar}$$

$$\psi(x) = \int_{-\infty}^{+\infty} \int_{0}^{+\infty} \Psi^n(X, P, \mathscr{b}) \, [(\boldsymbol{x} - X) \varphi_n(x, X, P, \mathscr{b})] \frac{dP d\mathscr{b}}{\pi \hbar \mathscr{b}}$$

$$\tilde{\psi}(p) = (-1)^n \int_{-\infty}^{+\infty} \int_{0}^{+\infty} \Psi^n(X, P, \mathscr{b}) \, [(\boldsymbol{p} - P) \tilde{\varphi}_n(p, X, P, \mathscr{b})] \frac{da dX}{\pi \hbar a}$$

It can be established that we have also the relations

$$\sum_{n=0}^{+\infty} |\Psi^n(X, P, \mathscr{b})|^2 = 1 \tag{21}$$

$$\int_{-\infty}^{+\infty} \int_{-\infty}^{+\infty} \frac{|\Psi^n(X, P, \mathscr{b})|^2}{2\pi\hbar} dX dP = 1 \tag{22}$$

In the framework of the probabilistic interpretation of quantum mechanics, these relations permits us to give the following physical interpretation of the phase space wavefunctions:

- The relations (17) and (21) allow to interpret the function $|\Psi^n(X, P, \mathscr{b})|^2$ as the probability to find a particle in a state $|n, X, P, \mathscr{b}\rangle$, for a fixed value of $X, P$ and $\mathscr{b}$ knowing that the state of this particle is $|\psi\rangle$.
- The relations (18) and (22) allow to interpret the function $\frac{|\Psi^n(X,P,\mathscr{b})|^2}{2\pi\hbar}$ as the density of probability to find a particle in a state $|n, X, P, \mathscr{b}\rangle$, for a fixed value of $n$ and $\mathscr{b}$, knowing that the state of this particle is $|\psi\rangle$.

We may list three particular examples of phase space wave functions by choosing respectively a particular value of the state vector $|\psi\rangle$

▪ For $|\psi\rangle = |x\rangle$

$$\Psi^n(X,P,\mathscr{b}) = \langle n,X,P,\mathscr{b}|x\rangle = \varphi_n^*(x,X,P,\mathscr{b})$$

▪ For $|\psi\rangle = |p\rangle$

$$\Psi^n(X,P,\mathscr{b}) = \langle n,X,P,\mathscr{b}|p\rangle = \tilde{\varphi}_n^*(p,X,P,\mathscr{b})$$

▪ For $|\psi\rangle = |n',X',P',\mathscr{b}'\rangle$

$$\Psi^n(X,P,\mathscr{b}) = \langle n,X,P,\mathscr{b}|n',X',P',\mathscr{b}'\rangle = \chi_{n'}^n(X,P,\mathscr{b},X',P',\mathscr{b}')$$

we may establish an explicit expression of the function $\chi_{n'}^n(X,P,\mathscr{b};X',P',\mathscr{b}')$ by performing, for instance, a calculation in the coordinate representation

$$\chi_{n'}^n(X,P,\mathscr{b};X',P',\mathscr{b}') = \int \varphi_n^*(x,X,P,\mathscr{b})\, \varphi_{n'}(x,X',P',\mathscr{b})dx \tag{23}$$

using the explicit expression of the function $\varphi_n(x,X,P,\mathscr{b})$ as given in the relation (3) and properties of Hermite polynomial, we can perform the calculation of the integral and find the explicit expression (see appendix)

$$\chi_{n'}^n(X,P,\mathscr{b};X',P',\mathscr{b}') = \mathcal{P}_{n'}^n(X,P,\mathscr{b},X',P',\mathscr{b}')e^{-\frac{(X-X')^2}{4(a^2+a'^2)}-\frac{(P-P')^2}{4(\mathscr{b}^2+\mathscr{b}'^2)}-\frac{i}{\hbar}\frac{(a'^2X+a^2X')}{a^2+a'^2}(P-P')} \tag{24}$$

in which $\mathcal{P}_{n'}^n(X,P,\mathscr{b},X',P',\mathscr{b}')$ is the polynomial

$$\mathcal{P}_{n'}^n(X,P,\mathscr{b},X',P',\mathscr{b}') = \sum_{l=0}^{n}\sum_{m=0}^{n'}[\frac{2(-1)^{n'-m}(i)^{l+m}\sqrt{n!\,n'!}\,(\mathscr{b}')^{n-l+m+\frac{1}{2}}(\mathscr{b})^{n'-m+l+\frac{1}{2}}}{l!\,(n-l)!\,m!\,(n'-m)!\,[2(\mathscr{b}^2+\mathscr{b}'^2)]^{\frac{n+n'+1}{2}}}$$

$$H_{n+n'-l-m}(\frac{X'-X}{\sqrt{2(a^2+a'^2)}})H_{l+m}(\frac{(P'-P)}{\sqrt{2(\mathscr{b}^2+\mathscr{b}'^2)}})] \tag{25}$$

we may consider three interesting particular cases:

- For $\mathscr{b} = \mathscr{b}'$(and $a = a'$), we have

$$\chi_{n'}^n(X,P,\mathscr{b};X',P',\mathscr{b}) = \mathcal{P}_{n'}^n(X,P,X',P',\mathscr{b})e^{-\frac{(X-X')^2}{8a^2}-\frac{(P-P')^2}{8\mathscr{b}^2}-i\frac{(X-X')(P-P')}{2\hbar}} \tag{26}$$

with

$$\mathcal{P}_{n'}^n(X,P,\mathscr{b};X',P',\mathscr{b}') = \sum_{l=0}^{n}\sum_{m=0}^{n'}[\frac{(-1)^{n'-m}(i)^{l+m}\sqrt{n!\,n'!}}{2^{n+n'}l!\,(n-l)!\,m!\,(n'-m)!}$$

$$H_{n+n'-l-m}(\frac{X'-X}{2a})H_{l+m}(\frac{P'-P}{2\mathscr{b}})] \tag{27}$$

- For $\mathscr{b} = \mathscr{b}'$(and $a = a'$) , $X = X'$ and $P = P'$ we have

$$\chi_{n'}^{n}(X,P,\mathscr{b};X,P,\mathscr{b}) = \langle n,X,P,\mathscr{b}|n',X,P,\mathscr{b}\rangle = \delta_{n'}^{n} = \begin{cases} 0 \; if \; n \neq n' \\ 1 \; if \; n = n' \end{cases}$$

The introduction of the function $\chi_{n'}^{n}(X,P,\mathscr{b};X',P',\mathscr{b}')$ permits to deduce interesting properties of the states $|n,X,P,\mathscr{b}\rangle$ and the phase space wavefunctions. In fact, as we have for any state $|\psi\rangle$

$$|\psi\rangle = \sum_{n} \Psi^{n}(X,P,\mathscr{b})\,|n,X,P,\mathscr{b}\rangle \tag{28}$$

It follows in particular that, for $|\psi\rangle = |n',X',P',\mathscr{b}'\rangle$, we have the relation

$$|n',X',P',\mathscr{b}'\rangle = \sum_{n} \chi_{n'}^{n}(X,P,\mathscr{b};X',P',\mathscr{b}')\,|n,X,P,\mathscr{b}\rangle \tag{29}$$

Let us now consider any state $|\psi\rangle$.It results from (28) that in the basis $\{|n,X,P,\mathscr{b}\rangle\}$ we have

$$|\psi\rangle = \sum_{n} \Psi^{n}(X,P,\mathscr{b})\,|n,X,P,\mathscr{b}\rangle \tag{30}$$

and in the basis $|n',X',P',\mathscr{b}'\rangle$

$$|\psi\rangle = \sum_{n'} \Psi^{n'}(X',P',\mathscr{b}')\,|n',X',P',\mathscr{b}'\rangle \tag{31}$$

Inserting the relation (29) in (31) and identifying with (30), we may deduce the interesting property which holds true for any phase space wavefunction $\Psi^{n}(X,P,\mathscr{b})$

$$\Psi^{n}(X,P,\mathscr{b}) = \sum_{n'} \chi_{n'}^{n}(X,P,\mathscr{b};X',P',\mathscr{b}')\Psi^{n'}(X',P',\mathscr{b}') \tag{32}$$

### 4-Differential equation satisfied by phase space wavefunctions

On one hand, from the relations (14) and (17), it can be deduced that for any phase space wavefunctions $\Psi^{n}(X,P,\mathscr{b}) = \langle n,X,P,\mathscr{b}|\psi\rangle$ we have the relation

$$\langle n,X,P,\mathscr{b}|\beth^{+}|\psi\rangle = \sum_{m} (2n+1\,)\mathcal{B}\delta_{m}^{n}\,\Psi^{m}(X,P,\mathscr{b}) = (2n+1)\mathcal{B}\Psi^{n}(X,P,\mathscr{b}) \tag{33}$$

and on the other hand, from the definition of differential operator representation, as given in our work [2], we have

$$\langle n,X,P,\mathscr{b}|\beth^{+}|\psi\rangle = \widetilde{\beth}^{+}\,\Psi^{n}(X,P,\mathscr{b}) \tag{34}$$

in which $\widetilde{\beth}^{+}$ is the differential operator representation of the momentum dispersion operator $\beth^{+}$ given in the relation (15).

$$\widetilde{\beth}^+ = -\frac{\hbar^2}{2\mathcal{B}}\frac{\partial^2}{\partial X^2} - 2\mathcal{B}\frac{\partial^2}{\partial P^2} - \frac{4i\mathcal{B}}{\hbar}X\frac{\partial}{\partial P} + \frac{2\mathcal{B}}{\hbar^2}X^2 \tag{35}$$

It follows from the relations (33), (34) and (35) that any phase space wavefunction $\Psi^n(X, P, \mathscr{z})$ satisfies the differential equation

$$[-\frac{\hbar^2}{2\mathcal{B}}\frac{\partial^2}{\partial X^2} - 2\mathcal{B}\frac{\partial^2}{\partial P^2} - \frac{4i\mathcal{B}}{\hbar}X\frac{\partial}{\partial P} + \frac{2\mathcal{B}}{\hbar^2}X^2]\Psi^n = (2n+1)\mathcal{B}\Psi^n \tag{36}$$

This equation is also the eigenvalue equation for the differential operator representation $\widetilde{\beth}^+$ of the momentum dispersion operator $\beth^+$ . And then, according to it, any phase space wavefunction $\Psi^n$ is an eigenfunction of $\widetilde{\beth}^+$ with the eigenvalue equal to $(2n+1)\mathcal{B}$.

The equation (36) can be also checked explicitly using the properties (32) of the phase space wavefunctions. In fact, according to this relation we have

$$\Psi^n(X, P, \mathscr{z}) = \sum_{n'} \chi^n_{n'}(X, P, \mathscr{z}; X', P', \mathscr{z}')\Psi^{n'}(X', P', \mathscr{z}')$$

so, taking into account the expression (35) of $\widetilde{\beth}^+$, we have

$$\widetilde{\beth}^+ \Psi^n(X, P, \mathscr{z}) = \sum_{n'} [\widetilde{\beth}^+ \chi^n_{n'}(X, P, \mathscr{z}; X', P', \mathscr{z}')]\Psi^{n'}(X', P', \mathscr{z}')$$

But according to the relation (23)

$$\chi^n_{n'}(X, P, \mathscr{z}; X', P', \mathscr{z}') = \langle n, X, P, \mathscr{z}|n', X', P', \mathscr{z}'\rangle = \int \varphi^*_n(x, X, P, \mathscr{z})\, \varphi_{n'}(x, X', P', \mathscr{z})dx$$

so, taking again into account the expression (35) of $\widetilde{\beth}^+$, we have

$$\widetilde{\beth}^+ \chi^n_{n'}(X, P, \mathscr{z}; X', P', \mathscr{z}') = \int [\widetilde{\beth}^+ \varphi^*_n(x, X, P, \mathscr{z})]\, \varphi_{n'}(x, X', P', \mathscr{z}')dx \tag{37}$$

Using the explicit expression of $\varphi_n$ given in (3) and the expression of $\widetilde{\beth}^+$ given in (35), we can establish after a long but straightforward calculation the relation

$$\widetilde{\beth}^+ \varphi^*_n(x, X, P, \mathscr{z}) = (2n+1)\mathcal{B}\varphi^*_n(x, X, P, \mathscr{z}) \tag{38}$$

This relation means that the function $\varphi^*_n(x, X, P, \mathscr{z})$ is an eigenfunction of $\widetilde{\beth}^+$ with the eigenvalue $(2n+1)\mathcal{B}$.

Using the relation (38), we can deduce from the relation (37) that

$$\widetilde{\beth}^+ \chi^n_{n'}(X, P, \mathscr{z}; X', P', \mathscr{z}') = (2n+1)\mathcal{B}\chi^n_{n'}(X, P, \mathscr{z}; X', P', \mathscr{z}') \tag{39}$$

Use of the relations (32), (35) and (39) allows to have, as expected, an explicit checking of (36).

**5- Conclusion**

In the phase space representation, the momentum dispersion operator $\beth^{+}$can be at the same time represented either with the diagonal matrix $\beth^{+n}_{m}$given in the relation (14) or with the differential operator $\overset{\frown}{\beth}^{+}$ given in the relation (15).This fact can be exploited to establish the equation (36) which is the eigenvalue equation of the operator $\overset{\frown}{\beth}^{+}$. According to this equation, any phase space wavefunction $\Psi^{n}(X,P,\mathscr{b}) = \langle n,X,P,\mathcal{B}|\psi\rangle$ is an eigenfunction of $\overset{\frown}{\beth}^{+}$ with eigenvalue equal to $(2n+1)\mathcal{B}$. This result can be considered as being logically expected since the elements of the basis $\{\,|n,X,P,\mathscr{b}\rangle\}$ defining the phase space representation themselves are the eigenstate of $\beth^{+}$.

In the section 3, we have done some recall concerning the phase space wavefunctions and give most of their main properties. Among these properties, we may notice a particular one which is given in the relation (32). This relation shows that the values of a phase space wavefunction $\Psi^{n}$ corresponding to different values of the parameters and variables $n, X, P$ and $\mathscr{b}$ can be linked using the function $\chi^{n}_{n'}(X,P,\mathscr{b};X',P',\mathscr{b}')$ defined in the relation (23). As discussed in the last part of the section 4, this particular property of phase space wavefunction given in the relation (32) permits also to perform an explicit checking of the eigenvalue equation (36) of $\overset{\frown}{\beth}^{+}$.

## Appendix

**Establishment of the expression on the function $\chi^{n}_{n'}(X,P,\mathscr{b};X',P',\mathscr{b}')$**

in the coordinate representation, we have

$$\chi^{n}_{n'}(X,P,\mathscr{b};X',P',\mathscr{b}') = \langle n,X,P,\mathscr{b}|n',X',P',\mathscr{b}'\rangle = \int \varphi^{*}_{n}(x,X,P,\mathscr{b})\,\varphi_{n'}(x,X',P',\mathscr{b})dx$$

Using the explicit expression of the function $\varphi_{n}$ given in the relation (3). We obtain

$$\chi^{n}_{n'}(X,P,\mathscr{b};X',P',\mathscr{b}') = [\frac{e^{\frac{a'^2X^2+a^2X'^2}{4a^2a'^2}}}{\sqrt{2^{n+n'+1}n!\,\pi a a'}}$$

$$\int_{-\infty}^{+\infty} H_{n}(\frac{x-X}{\sqrt{2}a})H_{n'}(\frac{x-X'}{\sqrt{2}a'})\, e^{-\{(\frac{a^2+a'^2}{4a^2a'^2})x^2-[\frac{a'^2X+a^2X'}{2a^2a'^2}-i\frac{(P-P')}{\hbar}]x\}}dx]$$

To perform the calculation of the integral, we introduce the generatrice function $G(x)$ of the Hermite Polynomials.

$$G(x,u) = e^{2xu-u^2} = \sum_{n=0}^{+\infty}\frac{u^{n}}{n!}H_{n}(x)$$

Using this function, we obtain the relation

$$\int_{-\infty}^{+\infty} e^{2(\frac{x-X}{\sqrt{2}a})u+2(\frac{x-X'}{\sqrt{2}a'})v-u^2-v^2}dx = \sum_{n=0}^{+\infty}\sum_{n'=0}^{+\infty}[\frac{u^{n}}{n!}\frac{v^{n'}}{n'!}\int_{-\infty}^{+\infty} H_{n}(\frac{x-X}{\sqrt{2}a})H_{n'}(\frac{x-X'}{\sqrt{2}a'})$$

$$e^{-\{(\frac{a^2+a'^2}{4a^2a'^2})x^2-[\frac{a'^2X+a^2X'}{2a^2a'^2}-i\frac{(P-P')}{\hbar}]x\}}dx]$$

The expression of the integral in the first member of this relation can be obtained easily using the relation

$$\int_{-\infty}^{+\infty} e^{-(Cx^2+Dx)}\,dx = \sqrt{\frac{\pi}{C}}\,e^{\frac{D^2}{4C}}$$

this obtained expression can be put in the form

$$\int_{-\infty}^{+\infty} e^{2(\frac{x-X}{\sqrt{2}a})u+2(\frac{x-X'}{\sqrt{2}a'})v-u^2-v^2}\,dx = \hbar\sqrt{\frac{\pi}{(\mathscr{b}^2+\mathscr{b}'^2)}}\,e^{-\frac{(X-X')^2}{4(a^2+a'^2)}-\frac{(P-P')^2}{4(\mathscr{b}^2+\mathscr{b}'^2)}-\frac{i}{\hbar}\frac{(a'^2X+a^2X')}{a^2+a'^2}(P-P')}F$$

in which $F$ is the function

$$F(X,X',P,P',\mathscr{b},\mathscr{b}'U,V) = e^{2(\frac{X'-X}{\sqrt{2(a^2+a'^2)}})U+2(\frac{P'-P}{\sqrt{2(\mathscr{b}^2+\mathscr{b}'^2)}})V-U^2-V^2}$$

$$= \sum_{k=0}^{+\infty}\sum_{q=0}^{+\infty}\frac{U^k}{k!}\frac{V^q}{q!}H_k(\frac{X'-X}{\sqrt{2(a^2+a'^2)}})H_q(\frac{P'-P}{\sqrt{2(\mathscr{b}^2+\mathscr{b}'^2)}})$$

$$= \sum_{k=0}^{+\infty}\sum_{q=0}^{+\infty}\frac{U^k}{k!}\frac{V^q}{q!}\frac{\partial^{k+q}F}{(\partial U)^k(\partial V)^q}\bigg|_{U=0,V=0}$$

with

$$\begin{cases} U = \dfrac{\mathscr{b}'}{\sqrt{\mathscr{b}^2+\mathscr{b}'^2}}u - \dfrac{\mathscr{b}}{\sqrt{\mathscr{b}^2+\mathscr{b}'^2}}v \\ V = \dfrac{i\mathscr{b}}{\sqrt{\mathscr{b}^2+\mathscr{b}'^2}}u + \dfrac{i\mathscr{b}'}{\sqrt{\mathscr{b}^2+\mathscr{b}'^2}}v \end{cases}$$

Using the relations between $(U,V)$ and $(u,v)$, we can establish the following relation

$$\frac{\partial^{n+n'}F}{\partial u^n\partial v^{n'}}\bigg|_{u=0,v=0} = \sum_{l=0}^{n}\sum_{m=0}^{n'}[\frac{(-1)^{n'-m}(i)^{l+m}n!\,n'!\,(\mathscr{b}')^{n-l+m}(\mathscr{b})^{n'-l+m}}{(\mathscr{b}^2+\mathscr{b}'^2)^{\frac{n+n'}{2}}}$$

$$\frac{\partial^{n+n'}F}{(\partial U)^{n+n'-l-m}(\partial V)^{l+m}}\bigg|_{U=0,V=0}]$$

$$= \sum_{l=0}^{n}\sum_{m=0}^{n'}[\frac{(-1)^{n'-q}(i)^{l+m}n!\,n'!\,(\mathscr{b}')^{n-l+m}(\mathscr{b})^{n'-l+m}}{l!\,(n-l)!\,m!\,(n'-m)!\,\left(\mathscr{b}^2+\mathscr{b}'^2\right)^{\frac{n+n'}{2}}}H_{n+n'-l-m}(\frac{X'-X}{\sqrt{2(a^2+a'^2)}})H_{l+m}(\frac{P'-P}{\sqrt{2(\mathscr{b}^2+\mathscr{b}'^2)}})$$

so we have for the expression of $F$ (as function of $u$ and $v$)

$$F = \sum_{n=0}^{+\infty}\sum_{n'=0}^{+\infty}\frac{u^n}{n!}\frac{v^{n'}}{n'!}\frac{\partial^{n+n'}F}{(\partial u)^n(\partial v)^{n'}}\bigg|_{u=0,v=0} = \sum_{n=0}^{+\infty}\sum_{n'=0}^{+\infty}\frac{u^n}{n!}\frac{v^{n'}}{n'!}\{\sum_{l=0}^{n}\sum_{m=0}^{n'}[\frac{(-1)^{n'-q}(i)^{l+m}n!\,n'!}{l!\,(n-l)!\,m!\,(n'-m)!}$$

$$\frac{(\mathscr{b}')^{n-l+m}(\mathscr{b})^{n'-l+m}}{\left(\mathscr{b}^2+\mathscr{b}'^2\right)^{\frac{n+n'}{2}}}H_{n+n'-l-m}(\frac{X'-X}{\sqrt{2(a^2+a'^2)}})H_{l+m}(\frac{P'-P}{\sqrt{2(\mathscr{b}^2+\mathscr{b}'^2)}})]\}$$

Introducing this expression (of $F$) in the following relation

$$\int_{-\infty}^{+\infty} e^{2(\frac{x-X}{\sqrt{2}a})u+2(\frac{x-X'}{\sqrt{2}a'})v-u^2-v^2}dx = \hbar\sqrt{\frac{\pi}{(\mathcal{b}^2+\mathcal{b}'^2)}}\, e^{-\frac{(X-X')^2}{4(a^2+a'^2)}-\frac{(P-P')^2}{4(\mathcal{b}^2+\mathcal{b}'^2)}-\frac{i}{\hbar}\frac{(a'^2X+a^2X')}{a^2+a'^2}(P-P')}F$$

$$= \sum_{n=0}^{+\infty}\sum_{n'=0}^{+\infty}[\frac{u^n}{n!}\frac{v^{n'}}{n'!}\int_{-\infty}^{+\infty} H_n(\frac{x-X}{\sqrt{2}a})H_{n'}(\frac{x-X'}{\sqrt{2}a'})\, e^{-\{(\frac{a^2+a'^2}{4a^2a'^2})x^2-[\frac{a'^2X+a^2X'}{2a^2a'^2}-i\frac{(P-P')}{\hbar}]x\}}dx]$$

we can obtain the expression of the integral containing product of Hermite polynomials and this result permits us to deduce that

$$\chi_{n'}^{n}(X,P,\mathcal{b};X',P',\mathcal{b}')$$

$$= [\frac{e^{\frac{a'^2X^2+a^2X'^2}{4a^2a'^2}}}{\sqrt{2^{n+n'+1}n!\,\pi a a'}}\int_{-\infty}^{+\infty} H_n(\frac{x-X}{\sqrt{2}a})H_{n'}(\frac{x-X'}{\sqrt{2}a'})\, e^{-\{(\frac{a^2+a'^2}{4a^2a'^2})x^2-[\frac{a'^2X+a^2X'}{2a^2a'^2}-i\frac{(P-P')}{\hbar}]x\}}dx]$$

$$= \mathcal{P}_{n'}^{n}(X,P,\mathcal{b},X',P',\mathcal{b}')e^{-\frac{(X-X')^2}{4(a^2+a'^2)}-\frac{(P-P')^2}{4(\mathcal{b}^2+\mathcal{b}'^2)}-\frac{i}{\hbar}\frac{(a'^2X+a^2X')}{a^2+a'^2}(P-P')}$$

in which $\mathcal{P}_{n'}^{n}(X,P,\mathcal{b},X',P',\mathcal{b}')$ is the polynomial

$$\mathcal{P}_{n'}^{n}(X,P,\mathcal{b},X',P',\mathcal{b}') = \sum_{l=0}^{n}\sum_{m=0}^{n'}[\frac{2(-1)^{n'-m}(i)^{l+m}\sqrt{n!\,n'!}\,(\mathcal{b}')^{n-l+m+\frac{1}{2}}(\mathcal{b})^{n'-m+l+\frac{1}{2}}}{l!\,(n-l)!\,m!\,(n'-m)!\,[2(\mathcal{b}^2+\mathcal{b}'^2)]^{\frac{n+n'+1}{2}}}$$

$$H_{n+n'-l-m}(\frac{X'-X}{\sqrt{2(a^2+a'^2)}})H_{l+m}(\frac{(P'-P)}{\sqrt{2(\mathcal{b}^2+\mathcal{b}'^2)}})]$$